\documentclass[12pt]{achemso}
\usepackage{graphicx} 
\usepackage{blindtext}
\usepackage{setspace}
\usepackage[utf8]{inputenc}
\usepackage{amsfonts, amsmath, amsthm, amssymb}
\usepackage{url}
\usepackage{multicol}
\usepackage{subcaption}
\usepackage{float}
\usepackage{titlesec} 
\usepackage{tikz}
\usepackage{transparent}
\usepackage{multirow}
\usepackage[toc,page]{appendix}
\usepackage{graphicx}
\usepackage{wrapfig}
\usepackage{caption}
\usepackage{color}
\usepackage{xcolor}
\usepackage{amsmath}
\usepackage{siunitx}
\usepackage{pdfpages}

\usepackage{geometry}

\title{Engineering deterministic, tunable, and reversible folds in graphene with the use  of ultrafast laser micro-patterned stretchable polymer substrate
}

\author{A.F.~Juarez Saborio}
\affiliation{Universite Claude Bernard Lyon 1, ILM Institut Lumiere Matiere, UMR CNRS 5306, 69622 Villeurbanne cedex, FRANCE}
\alsoaffiliation{Universite Jean Monnet, Laboratoire Hubert Curien, UMR CNRS 5516, 42000 Saint-Etienne, FRANCE}
\author{F.~Bourquard}
\affiliation{Universite Jean Monnet, Laboratoire Hubert Curien, UMR CNRS 5516, 42000 Saint-Etienne, FRANCE}
\author{R.~Galafassi}
\affiliation{Universite Claude Bernard Lyon 1, ILM Institut Lumiere Matiere, UMR CNRS 5306, 69622 Villeurbanne cedex, FRANCE}
\author{A.~Claudel} 
\author{L.~Marty} 
\affiliation{Univ. Grenoble Alpes, CNRS, Grenoble INP, Institut Néel, 38000 Grenoble, France}
\author{A.~Piednoir} 
\author{M.~Mercury} 
\author{R.~Fulcrand}
\author{C.~Albin} 
\affiliation{Universite Claude Bernard Lyon 1, ILM Institut Lumiere Matiere, UMR CNRS 5306, 69622 Villeurbanne cedex, FRANCE}
\author{V.~Barnier} 
\affiliation{Mines Saint-Etienne, LGF Laboratoire Georges Friedel, UMR CNRS 5307, F-42023 Saint-Étienne cedex 2, FRANCE}
\author{F.~Garrelie} 
\affiliation{Universite Jean Monnet, Laboratoire Hubert Curien, UMR CNRS 5516, 42000 Saint-Etienne, FRANCE}
\author{A.~San-Miguel}
\author{F.~Vialla}
\affiliation{Universite Claude Bernard Lyon 1, ILM Institut Lumiere Matiere, UMR CNRS 5306, 69622 Villeurbanne cedex, FRANCE}
\email{fabien.vialla@univ-lyon1.fr}

\date{2025}

\begin{document}

\maketitle
\doublespacing

\newpage

\begin{abstract}

The unique atomic monolayer structure of graphene gives rise to a broad range of remarkable mechanical folding properties. However, significant challenges remain in effectively harnessing them in a controllable and scalable manner. In this study, we introduce an innovative approach that employs micron-scale cavities, fabricated through ultrafast laser patterning, in a stretchable polymer substrate to locally modulate adhesion and strain transfer to a graphene monolayer. This technique enables the deterministic induction of single folds in graphene with fold dimensions, width and height in the hundreds of nanometers, tunable through the geometry of the polymer cavities and the applied strain. Importantly, these folds are reversible, returning to a flat morphology with minimal structural damage, as confirmed by Raman spectroscopy. Additionally, our method allows for the creation of fields of folds with reproducible periodicity, defining clear potential for practical applications. These findings pave the way for the development of advanced devices that would leverage the strain and morphology-sensitive properties of graphene.
    
\end{abstract}

\section*{Introduction}

Graphene presents a unique one-atom-thick structure making it a model system for investigating the mechanics of thin layers at the ultimate two-dimensional limit~\cite{AKINWANDE201742,sun2021APR,galiotis2015ARCBE}. It exhibits inherent non-deterministic local folds and more widespread undulations, with a large variety of morphologies reported and classified as ripples, wrinkles, crumples, or folds~\cite{deng2016wrinkled,Megra_2019}. Graphene's folding behavior has attracted substantial scientific interest, ranging from its similarities to macroscale textiles or sheets~\cite{LOPEZPOLIN202218,vandeparre2010wrinkling}, to the unique characteristics driven by its atomistic thinness and the nature of its bonds~\cite{ZhangPhysRevLett.106.255503,Zhu2012nanolett}.
Interestingly, folding induces local strains with the potential to modify the exceptional properties of graphene in meaningful ways, including electronic, optical, mechanical, and thermal attributes~\cite{deng2016wrinkled,peng2020strain}. This tunability includes the opening of a bandgap~\cite{ni2008uniaxial}, the enhancement of mobility of charge carriers~\cite{Zhu2012nanolett,Carrillo-Bastos2014}, the alteration of the interface properties, with different folding morphologies promoting covalent over van der Waals bonding~\cite{Silva2015}, or hydrophilicity over hydrophobicity~\cite{zang2013multifunctionality}. Transitioning from naturally occurring curvatures to deterministic, controllable folds appears crucial for harnessing the full potential of graphene in functional device applications.

Various methods relying on one- or two-dimensional mechanical strain, through the use of thermal treatment~\cite{natBao2009contripp,tapaszto2012breakdown}, control of environmental pressure~\cite{metten2014all,amaral2021delamination}, or compliance to a stretchable substrate~\cite{zang2013multifunctionality,wang2011super,TsoukleriSMALL2009,FrankACS2010,jiang2014interfacial}, have successfully created stochastic crumples, wrinkles, and ripples in initially pristine graphene layers. Recent advances in device design have led to the demonstration of reversible tuning of these stochastic fields of folds in graphene~\cite{Leem_2019}. Alternatively, more deterministic single or multiple folded graphene structures can be engineered during the transfer process of the graphene sheet, influenced by the patterning of either the transferring stamp~\cite{hallam2015controlled} or the underlying substrate~\cite{reserbat2014strain,pacakova2017mastering}. However, it is important to note that these structures remain static once formed.
Achieving both deterministic and tunable folds has been demonstrated only at a local scale using advanced tools. This usually involves the actuation with a microscopic tip~\cite{chen2019atomically,chang2018small}, which can result in reversible single layer folding, as well as complex manipulation following patterned cutting~\cite{blees2015graphenekiri}. These achievements are of high importance as proof of principle for innovative concepts related to origami and kirigami~\cite{ZHANG20211MSERR}. However, challenges related to scalability remain, underscoring the need for further research to enable the practical application of these techniques in larger-scale systems towards functional devices.

This work presents innovative experimental methods aimed at achieving practical control over graphene folded structures. Specifically, we employ controlled uniaxial actuation of graphene deposited on a polymer substrate which has been pre-stretched and textured with micron-scale cavities using ultrafast laser techniques. The graphene morphology is characterized through atomic force microscopy (AFM) coupled with structural optical Raman spectroscopy.  Our findings demonstrate that large compression can induce localized folding in graphene with the folding features directly correlated to the pre-textured patterns. This capability allows for precise control over the location and height of the folds. The formation of larger scale, periodic folded patterns is discussed. The demonstration of deterministic, tunable, and reversible folding in graphene establishes a reliable foundation for the fabrication of functional devices, based on more complex designs taking advantage of the unique morphology-dependent properties of graphene.

\section*{Results and Discussions}

\subsection*{Scientific approach and sample design}

\begin{figure}[t]
\centering
\includegraphics[width=.67\linewidth]{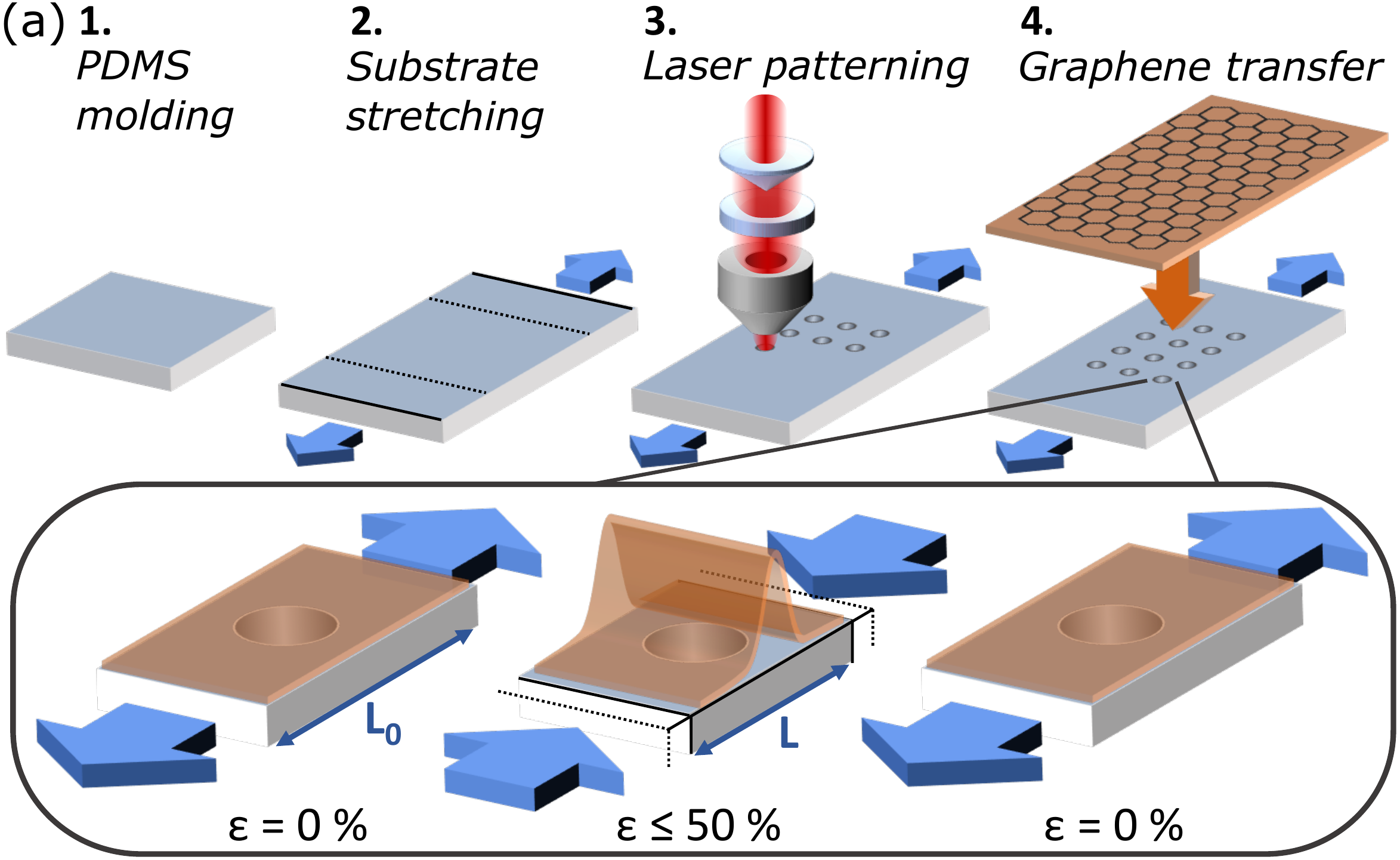}
\includegraphics[width=.32\linewidth]{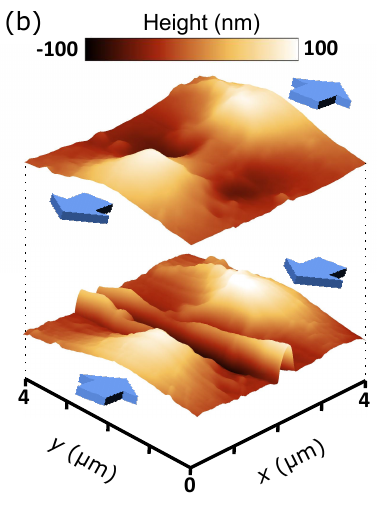}
\caption{
(a) Schematic representation of the four-step preparation of the PDMS substrate (light blue) molding, stretching, patterning, and the graphene (orange) transfer. The inset presents the deterministic, reversible folding of graphene at a patterned cavity through the application of a cycled uniaxial strain (blue arrows) to the substrate. (b) Experimental realization illustrated with AFM height maps of graphene on a micron-scale cavity-patterned PDMS substrate for $\epsilon = 0$ (top) and $\epsilon = 50 \%$ (bottom).
}
\label{fig:strategy}
\end{figure}

Our experimental strategy involves the application of extreme uniaxial compressive stress to graphene, achieved through the relaxation of a previously stretched substrate that incorporates localized heterogeneities following a specific patterned design, as schematized in Fig.~\ref{fig:strategy}a. This approach is informed by previous studies demonstrating the limited strain transferred to graphene when subjected to strong substrate compression, which can lead to phenomena such as delamination, buckling and wrinkling upon reaching a critical strain threshold~\cite{TsoukleriSMALL2009,FrankACS2010,jiang2014interfacial,Megra_2019}. These structures are mostly stochastic, with the adhesion properties between the graphene and the substrate playing a crucial role. Therefore, we propose an engineered solution that utilizes graphene suspended regions at the micro-scale obtained through substrate patterning to induce the controlled localization of such folded structures. To implement this, we constructed a custom traction machine designed to clamp a thin polymer stripe (8 x 18 x 0.75 mm$^3$ in size) and apply precise uniaxial strain through controlled stretching and unstretching with a resolution down to 10 micrometers (see Supplementary Material, SM, section 1). Polydimethylsiloxane (PDMS) was selected as the polymer substrate due to its favorable adhesion properties with graphene, as well as its moldability, stretchability, and reliability for repeated, precise stretching cycles. We could reach an initial tensile strain of 100~\% without material failure. Cavities were etched into the stretched PDMS substrate using ultrafast laser techniques, which offer rapid, versatile, and large-surface processing capabilities. The experimental setup employed a 1030 nm PHAROS laser focused in a Bessel beam, allowing for energy injection into the substrate's depth rather than merely affecting the surface~\cite{khonina2020bessel,bhuyan2010ultrafast}. Despite the technical challenges posed by the transparency~\cite{mao2004dynamics,gattass2008femtosecond} and the low melting point of PDMS~\cite{rebollar2011assessment,rebollar2015laser,stankova2016optical}, and through meticulous optimization of laser parameters including pulse duration and energy, we achieved intricate patterning with high spatial resolution, down to 1 micron. Various matrices of dots were etched into several samples to ensure redundancy and test different configurations. After the PDMS substrate's stretching and patterning steps, graphene grown by chemical vapor deposition (CVD) was transferred onto the stretched PDMS using a wet transfer method~\cite{HanCVDafm2014,coraux2013functional,arjmandi2018large}. The achieved relatively large and homogeneous coverage of graphene allowed for the comparison of fold formation in several etched cavities and enabled the investigation of potential collective effects. To quantify the percentage of compression applied to the graphene layer, we define an adapted strain parameter $\epsilon = (L_0 - L) / L_0$, where $L_0$ and $L$ are the initial length of the stretched PDMS substrate when graphene is transferred and the current length of PDMS, respectively, see Fig.~\ref{fig:strategy}a. A value of zero corresponds to a fully stretched substrate with no strain induced in the transferred graphene, and positive values indicate a release of the substrate, resulting in uniaxial compressive strain in the graphene. This parameter allows for a systematic investigation of the relationship between applied macroscopic strain and the resulting morphological changes in graphene. The response of graphene to the applied strain was primarily characterized using AFM, yielding spatial topology maps which demonstrate the nucleation of a deterministic fold as intended (see Fig.~\ref{fig:strategy}b), with quantitative evaluation of its geometrical features. Raman spectroscopy was used as a complementary technique to provide additional insight into the material's properties~\cite{ferrari2006raman,malard2009raman}. 
Further details of these experimental procedures and sample designs are elaborated in the Methods section.

\subsection*{Deterministic and tunable single folds}

\begin{figure}[t]
\includegraphics[width=.51\textwidth]{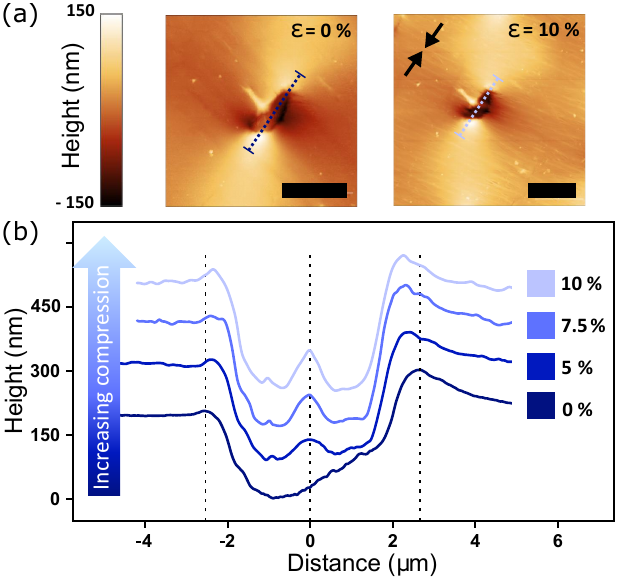} 
\caption{ 
(a) AFM height maps of graphene on a 5.2~$\mu$m cavity-patterned PDMS substrate for applied compression $\epsilon = 0$ (left) and $10 \%$ (right). Scale bars correspond to 5 $\mu$m. Black arrows indicate the axis of compression. (b) AFM height linecuts along the compression axis over the cavity suspended graphene (see dotted lines in a) for several compression $\epsilon$. Curves are offset with increasing $\epsilon$ for clarity. Dotted vertical lines highlight the progressive formation of the fold and reduction of the cavity size.
}
\label{fig:smallepsilon}
\end{figure}

We first focus on the behavior of a single large patterned cavity under low compressive constraints $\epsilon \leq 10\%$. The primary objective is to validate the occurrence of deterministic nucleation of a fold within the suspended graphene region. In Fig.~\ref{fig:smallepsilon}, the AFM height profile of a line cut through a covered 5.2~$\mu$m cavity reveals the response of the graphene layer to varying strain. At $\epsilon = 0\%$, the as-transferred graphene layer slightly conforms to the contours and is suspended over the cavity. During compression, the distance between the walls narrows, demonstrating effective, yet not complete, local transmission of the globally applied strain. We note that the roughness of the supported regions of graphene increases without the formation of distinct folds, as expected for a graphene layer deposited on an unpatterned compressed polymer. In contrast, within the suspended region of the cavity, we observe the emergence of one distinct single fold. With increasing compression, the fold progressively rises, correlating with the levels of applied compression. These observations clearly demonstrate the deterministically localized formation of a fold within a predetermined cavity, with the fold initially remaining confined within the cavity under this low constraint regime. 

A complete description of the folding mechanisms remains beyond the scope of our experimental demonstration. 
In particular, we cannot discriminate between a direct nucleation of the fold inside the cavity and a selection and amplification of a single small stochastic fold coming from the side.
However, we estimate that the observed behavior is primarily enabled by the initial shape of the cavity edges, which exhibits a distortion in symmetry due to the initial uniaxial stretching of the polymer, with edges rising along this axis and falling in the orthogonal direction (see SM section 2). The suspended graphene sheet, by slightly conforming to this shape (Fig.~\ref{fig:strategy}b), acquires an anisotropic morphology which may help the nucleation, and further rise in height, of the fold.
More specifically, the stretching due to Poisson effect along the axis perpendicular to the compression may play a crucial role in preventing the sheet from falling into the cavity, and instead favoring an upward folding behavior. To support this claim, we computed and evaluated the competing stretching and bending energies~\cite{yamamoto2012princess} in simplified cases (see SM section 3), suggesting an energetically more favorable configuration for the fold extending upwards in the cavity.
Additionally, we find a slight deviation in the fold profile compared to the reported stochastic folds in thin films~\cite{vella2009macroscopic} and 2D materials layers~\cite{brennanAMI2015}, which advocates for further investigation in the involved folding mechanism (see SM section 4).

\begin{figure}[p!]
\centering
\includegraphics[width=.98\textwidth]{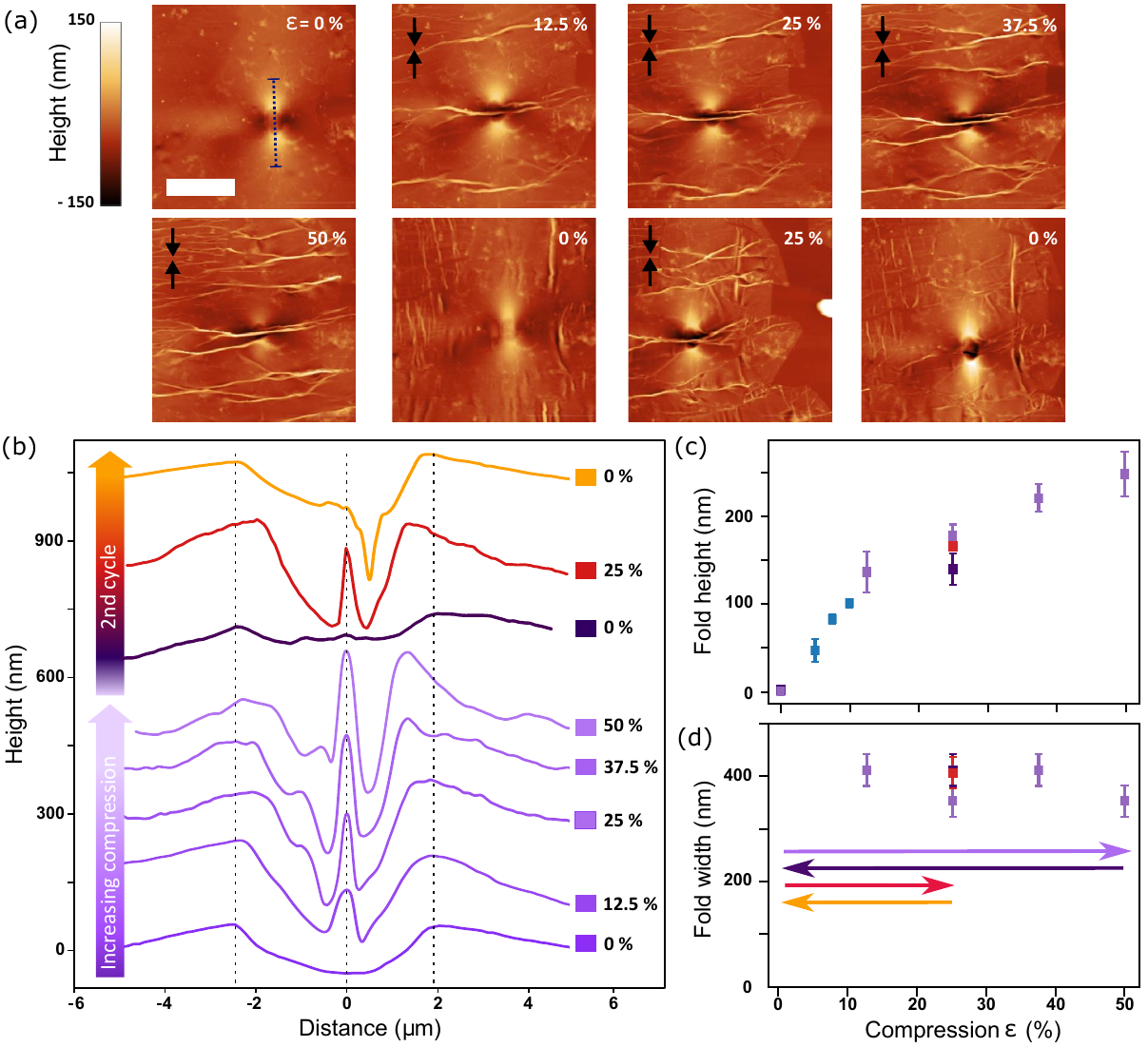} 
\caption{
(a) AFM height maps of graphene on a 2.4~$\mu$m cavity-patterned PDMS substrate for a uniaxial strain application cycle, from top left to bottom right, with compression $\epsilon$ indicated in white. The scale bar corresponds to 5 $\mu$m. Black arrows indicate the axis of compression. (b) AFM height linecuts along the compression axis over the graphene suspended on the cavity (see dotted line in a) for several compression $\epsilon$. Curves are offset following the cycle process. Dotted vertical lines highlight the progressive and reversible formation of the fold and reduction of the cavity size. Extracted fold height (c) and width (d) as function of compressive strain $\epsilon$. Square data points correspond to the presented cycled folding with first compression (purple), first release (black), second compression (red), and final release (yellow) highlighted with horizontal arrows. Blue points correspond to data presented in Fig.~\ref{fig:smallepsilon}.
}
\label{fig:largeepsilon}
\end{figure}

To further investigate the response of graphene to varying levels of compression, we repeated the process with different samples subjected to higher constraints, achieving a maximum compression percentage of $\epsilon = 50\%$. The AFM profiles for another typical single fold within a cavity are presented in Fig.~\ref{fig:largeepsilon}a,b. In Fig.~\ref{fig:largeepsilon}c, we observe that the height of the nucleated fold, defined as the height distance between the lowest and highest points in the linecut, reaches at low strain a similar value to the one presented above, illustrating the reproducibility of our approach. The fold height further increases to 300~nm under the highest applied compressive strain, which is significantly greater than typical values observed for stochastic wrinkles induced by compressive substrates. Interestingly, a monotonously increasing trend, close to linearity, is observed within this range, highlighting the potential for precise control over fold dimensions through systematic adjustments of strain levels, depending on the desired height. The ability to finely tune the height of graphene folds represents a highly sought-after capability which is here demonstrated. Additionally, the full width evolution of the fold, defined as the lateral distance between the two minima shown in Fig.~\ref{fig:largeepsilon}d, remains relatively constant at around 300~nm for a cavity size of 2.4~µm prior to compression. In comparison, for a cavity size of 5.2~µm as presented in Fig.~\ref{fig:smallepsilon}, the full width is also constant but measures 1200~nm. This suggests that the full width of the fold is primarily driven by the size of the cavity, whereas the fold height appears to be more universal and influenced by the applied strain. 

Noteworthy, we observe in Fig.~\ref{fig:largeepsilon}a that, for these higher values of compression, the fold surpasses the depth of the cavity within which it was initially confined, extending beyond its boundaries in the direction orthogonal to the applied strain. In addition to this main fold, secondary folds emerge as compression increases, becoming more pronounced and contrasting as the level of compression rises. The characterization of this fold field will be discussed in a final section.

\subsection*{Reversibility and cycles}

The presented sample has been further tested in a cyclic process, following the maximum compression of $\epsilon = 50\%$. A re-stretching step returned the sample to its original state at $\epsilon = 0\%$. Then a second cycle was initiated, applying compression again before returning to the initial unstrained configuration. This process establishes a cycle of folding and unfolding. AFM mapping and linecut profiles in Fig.~\ref{fig:largeepsilon}a,b reveal that the folding process can be effectively reversed, allowing the graphene to return to a state where the prominent fold observed in the cavity is no longer present. 
Similarly, the secondary folds that formed during compression also disappear upon reversing the strain. 
Importantly, we quantitatively measure the same full width and fold height for a given strain of $\epsilon = 25\%$ across the configurations of the first compression, first return, and second compression. 
This quantitative reproducibility is essential for the development of tunable devices. 
We further evaluated the evolution of the total length of the suspended graphene over the cycle (see SM section 5). We observe that, after an alternation of sticking and slipping in the first compression, a clamped behavior is reached at large compression and kept over the different steps of the reversible cycle. The reaching of such stabilized configuration is of strong interest towards reliable device performances. 

However, we note a progressive degradation of the integrity of the graphene layer, culminating in a rupture within the cavity during the final step of the second cycle.
Folds and cracks along the strain axis, perpendicular to the studied fold, are observed when returning to $\epsilon = 0\%$ macroscopic configuration. This phenomenon may correspond to strain in the perpendicular direction, potentially due to excessive stretching during the return cycle. A meticulous investigation of the folding cycles is warranted to evaluate these effects, optimize integrity over multiple cycles, and validate the applicability of this approach in device contexts. 
Finally, a characterization of the folding behavior in other graphene-covered cavities is presented in the SM section 6, further substantiating our conclusions.

\begin{figure}[t]
\begin{center}
\includegraphics[width=0.95\textwidth]{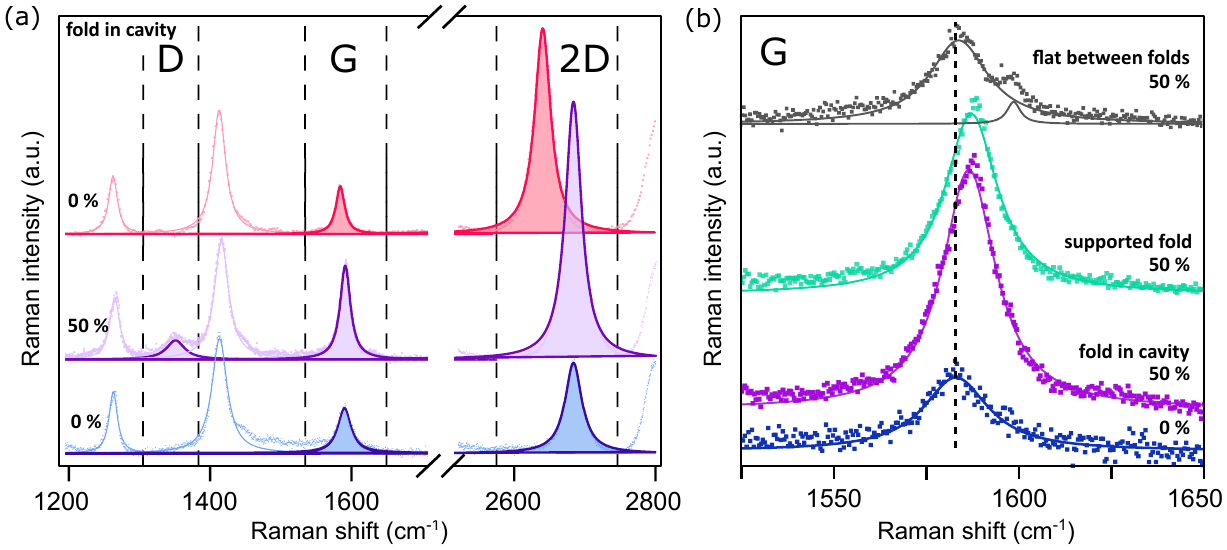} 
\caption{
(a) Raman spectra taken at the cavity position in the sample presented in Fig.~\ref{fig:largeepsilon}, for as-deposited ($\epsilon = 0\%$, blue), first compression ($\epsilon = 50\%$, purple), and first release ($\epsilon = 0\%$, red) configurations. Raman signals are normalized on the PDMS peaks. Vertical dashed lines defines the spectral regions of interest for D, G and 2D-bands in graphene. Other features arise from the PDMS substrate. Dots represent data, and lines the Lorentzian fitted curves with shades to highlight the graphene peaks. (b) Close-up of graphene Raman G-band at the same cavity position for as-deposited $\epsilon = 0\%$ (blue) and compressed $\epsilon = 50\%$ (purple) configurations. Spectra for compressed graphene on a secondary fold outside the cavity (green) and on a flat region between folds (grey) are presented. The vertical dashed line highlights the shifts induced by compression and folding. Dots represent data, and lines the Lorentzian fitted single G or splitted G$_-$/G$_+$ peaks.
All curves are offset and labeled.
}
\label{fig:raman}
\end{center}
\end{figure}

While AFM images provide valuable insight into the morphological modifications experienced by the graphene layer, additional characterization with Raman spectroscopy provides structural and electronic information, including its crystalline quality, defect density, doping level, and strain, while being fast and non-destructive~\cite{ferrari2006raman,malard2009raman}.
We present in Fig.~\ref{fig:raman}a the results of Raman measurements conducted on the graphene suspended over the cavities subjected to varying levels of applied strain. Given the dimensions of the cavities ($\gtrsim 2$ $\mu \text{m}$), the diameter of the laser spot (700 nm), and the width of the one-dimensional fold (300 nm), the Raman signal predominantly originates from the part of the graphene experiencing the progressive curvature as a result of the folding process.
In the uncompressed state, the presence of the G-band at 1581~cm$^{-1}$ and the 2D-band at 2684~cm$^{-1}$, along with a 2D/G intensity ratio of 4.2, indicates that the graphene is a pristine monolayer with minimal strain~\cite{ferrari2006raman,malard2009raman}. The absence of the D-band, typically observed around 1350 cm$^{-1}$, further corroborates the integrity of the deposited graphene layer, suggesting a low density of defects.

Upon compressing and folding of the graphene within the cavity, we observe the emergence of the D-band in the Raman signal. This phenomenon can be attributed to the increased curvature in the fold, which induces local strain and disrupts the planar symmetry of the graphene lattice, thereby activating D-band scattering processes~\cite{picheau2021intense}. Alternatively, this activation may be associated with the formation of irreversible structural defects within the graphene lattice~\cite{ferrari2006raman,malard2009raman}. When the strain is released and the graphene returns to its stretched, unfolded state, the D-band disappears, leaving only the G and 2D-bands visible. The absence of the D-band upon unfolding implies that the symmetry-breaking effects are largely reversible, indicating that the folding-unfolding process does not induce permanent defects or bond breaking in the graphene lattice. Instead, it results in temporary curvature changes that can be restored upon unfolding~\cite{GALAFASSIcarbon2024}.

During the folding process, we also observe a blue shift up to 5.7~cm$^{-1}$ in the G-band (Fig.~\ref{fig:raman}b), as expected for an increase in bond strength due to compression or folding. When interpreted as uniaxial strain, the value would correspond to a small residual strain of approximately 0.2\%~\cite{FrankACS2010}, indicating that most of the stress is converted in terms of bending of the graphene. Upon unfolding, the G-band shift returns to its original value, with no significant broadening of the line (Fig.~\ref{fig:raman}a), further indicating a preservation of the graphene structural integrity in the process. 
Interestingly, we observe for the 2D-band a small blueshift when folded, and a significant redshift when stress is released, both accompanied by an increase in the 2D/G intensity ratio. This observation does not align with a simple reversible compression or bending of the layer. It thus suggests that, although the graphene layer is restored to its flat configuration, modifications, possibly in the electronic structure, persist as a result of the previous deformation. These changes call for further investigation to fully understand their origins.
Overall, the morphological and structural characterizations of the behavior of suspended graphene upon a cycle of uniaxial compression validate the deterministic, reversible, and tunable formation of single folds with unprecedented heights. 

\subsection*{Patterns and periodicity}

\begin{figure}[t]
\begin{center}
\includegraphics[width=.98\linewidth]{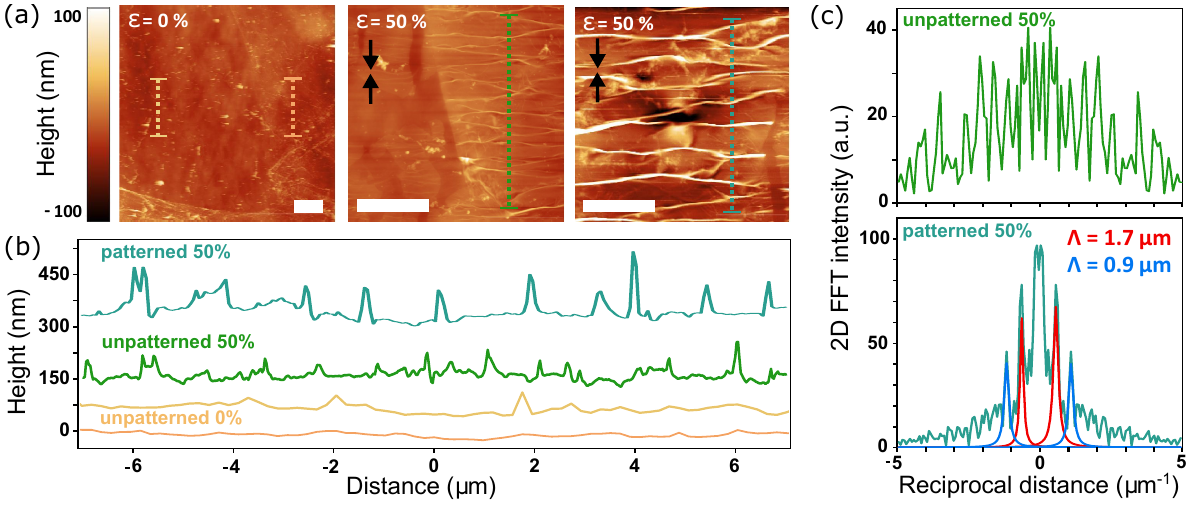}
\caption{
(a) AFM height maps of graphene on a pristine PDMS substrate prior (left) and after (center) uniaxial compression of $\epsilon = 50 \%$, and on a different patterned cavity at $\epsilon = 50 \%$ (right) of the sample in Fig.\ref{fig:largeepsilon}. Scale bars correspond to 5 $\mu$m. Black arrows indicate the axis of compression. (b) AFM height linecuts along the dotted lines with respective colors. Curves are offset and labeled. (c) 2D FFT linecuts of AFM data in (a) along the compression axis for the unpatterned (top) and patterned (bottom) configurations at $\epsilon = 50 \%$. Fitted spatial periodicity values are indicated as $\Lambda$ terms.}
\label{fig:largescale}
\end{center}
\end{figure}

In the previous sections, we have shown that, for sufficiently large values of $\epsilon$, the fold emerging from a textured hole can extend outwards for several microns, in a direction orthogonal to the compression (see Fig.~\ref{fig:largeepsilon}). 
This is accompanied by the formation of similar parallel folds distributed along the direction of the compression.
To investigate how substrate texturing influences fold formation and potential emergence of patterns and periodicity on a larger scale, we compared pre-textured with non-textured samples. 
Here we only investigated texturation with large distances of 15~microns between cavities (see SM section 1).
In Fig.~\ref{fig:largescale}, AFM maps and height profiles in several configurations reveal the clear effect of the PDMS texturation on the morphology of the folds compared to stochastic ones, induced either during the transfer of graphene or by the compression on pristine PDMS. 
Root mean square (RMS) roughness quantitatively reflects this observation when comparing the different configurations over the same area. We find values of 3~nm for stretched non-textured PDMS, 7~nm after graphene deposition, and 20~nm after compression. A notably higher RMS roughness of 38~nm is observed for the presented compressed graphene on pre-textured PDMS.
Moreover, in the case of graphene on non-textured PDMS, the folds show a consistent directionality orthogonal to the compression, yet with variable heights and no clear spatial organization. In contrast, pre-textured PDMS appears to induce a more regular, periodic folding pattern of folds, with heights that are consistently greater, often exceeding 100~nm for $\epsilon = 50 \%$, and separated by a distance of around 1.5~microns.

To obtain a more rigorous analysis of fold spacing, we employed a 2D Fast Fourier Transform (FFT) approach as presented in Fig.~\ref{fig:largescale}c. The 2D FFT offers a frequency-based analysis across the entire sample image, providing a robust, comprehensive view of fold periodicity, which is more reliable than isolated height profiles or roughness metrics.
Notably, the spatial FFT along the compression axis of the pre-textured, compressed sample shows distinct peaks, signaling a periodic structure. The dominant peak corresponds to a wavelength of approximately 1.7~micron, indicative of the primary fold pattern periodicity. Additionally, a secondary peak appears at around 0.9~micron, which could represent a harmonic of the primary periodicity or a smaller-scale periodic feature. This may relates to the observed splitting of folds in Fig.~\ref{fig:largescale}a, corresponding to a universal wrinkling hierarchy behavior already demonstrated in graphene and other thin sheets~\cite{vandeparre2010wrinkling}. 
This procedure was applied to the folds presented in Fig.~\ref{fig:largeepsilon} with similar quantitative conclusion (see SM section 7).
The observed clear periodic structure differs markedly from the diffuse peaks in the untextured sample’s FFT, highlighting that the periodicity arises specifically from the pre-texturing pattern rather than random folding.

We expanded the Raman spectroscopy study in Fig.~\ref{fig:raman}b by examining secondary folds located outside the cavities, as well as flat regions where the graphene remains compressed but unfolded. 
In the former case, the G-band presents the same profile as the one measured for the fold inside the cavity, revealing a similar nature in the strain and deformation in the layer.
In the latter case, Raman spectra reveal both a blueshift and a splitting of the G-band into two components, G$_-$ and G$_+$, characteristic of uniaxial or anisotropic strain, where compressive stress is applied along a specific direction~\cite{FrankACS2010}. We extract a G-band splitting of 15 cm$^{-1}$ at $\epsilon = 50 \%$, which corresponds to an estimated uniaxial strain of 0.8\%. We find that the observed strain in graphene is far from the macroscopic strain applied during compression, which was 50\%. Yet, it directly aligns with the maximum reported Raman strain value for flat graphene on a compressed polymer, before behaviors like slipping, delamination, or buckling occur~\cite{TsoukleriSMALL2009,FrankACS2010,jiang2014interfacial}. As shown in Fig.~\ref{fig:largescale}a, these patterns would emerge in a stochastic manner for a pristine polymer. In contrast, here we find that the stress is first relaxed through the bending of the layer at the controlled nucleation localization inside the cavity. Coming back to Fig.~\ref{fig:largeepsilon}a, as the compression is increased, the cavity-nucleated fold expands outside along the direction orthogonal to the compression, and is followed by the formation of parallel satellite folds rising in a progressive, symmetric manner, leading to the periodic pattern at high $\epsilon$ values already commented in previous sections and SM section 7. Notably, this directly aligns with the behavior of a macroscopic sheet under uniaxial strain~\cite{vella2009macroscopic}. 
We infer that the cavity-nucleated fold drives the competition between the stretching, adhesion, and bending energies inside the graphene sheet throughout the full compression process. This first deterministically localized fold formation in the cavity defines regions at fixed distances from it where the high compressive stress is preferentially and sequentially relaxed by delamination of a wrinkled fold leading to the appearance of the observed patterns. The exact parameters governing the extracted quantitative features, such as the spatial periodicity, height and width of the folds, or the strain thresholds for satellite fold formation, remain to be investigated. 
In particular, the role of the distance between the cavities, fixed at a long value of 15~microns in our study, should be tested. Specifically, varying the spacing, especially towards lower values in the range of observed periodicity, may show the emergence of collective effects in the folding pattern~\cite{pacakova2017mastering}.
From this understanding, more advanced designs can be developed with the aim of engineering complex large-scale patterns while conserving the demonstrated tunability and reversibility.

\section*{Conclusion}

In conclusion, we developed a novel experimental approach to achieve deterministic folding of graphene based on precise ultrafast laser patterning of the polymer substrate used to apply uniaxial compressive stress. We demonstrated the nucleation of a fold in graphene suspended over the patterned cavity, with tunable properties governed by the cavity size and applied compressive strain. Reversibility between the flat configuration and several hundred nanometer-high folds was achieved. The formation of spatially periodic patterns of folds was described and discussed, with strong applicative potential, yet calling for further investigation. 
This novel approach defines clear guidelines for the engineering of tunable graphene devices through its morphology-sensitive properties. To this end, efforts on material integrity over several runs are still requested. The development of advanced designs of cavity patterns in polymer is highly sought-after to create deterministic, tunable, and reversible large fields of graphene folds with the corresponding emerging functional features in the field of optics, electronics, or surface reactivity~\cite{deng2016wrinkled,Megra_2019,kim2019uniaxially}. This work may also be adapted and generalized to other 2D materials towards a larger range of tunable engineered properties, such as strain-dependent (single photon) emitters in transition metal dichalcogenides~\cite{castellanos2013local,kumar2015strain}.

\section*{Experimental Methods}

\textbf{PDMS substrate.} Polydimethylsiloxane (PDMS) substrates were prepared by mixing Sylgard 184~base and curing agent in a 10:1 ratio. It was then degassed under vacuum, and cured at 80~°C for 2~hours. The cured PDMS was cut into 18~mm × 8~mm rectangular strips and uniaxially stretched, held fixed using the traction machine presented in SM.

\textbf{Ultrafast laser patterning.} Laser texturing was performed using a Yb:KGW ultrafast laser system (1030 nm central wavelength, 2 ps pulse duration, 10 kHz repetition rate). The average laser power was 107 mW, corresponding to a pulse energy of 10.7 $\mu$J and a fluence of $\sim$1.51 J/cm$^2$, with a beam waist radius of $\sim$15 $\mu$m. Patterns were generated as square matrices of 15 $\mu$m × 15 $\mu$m
cavities with 15 $\mu$m spacing between features and an average depth of $\sim$2 $\mu$m, as measured by AFM.

\textbf{Graphene growth and transfer.} Monolayer graphene films were synthesized by chemical vapor deposition (CVD) on copper foils and transferred onto the laser-textured PDMS. The copper substrate was etched by immersing the PDMS/graphene/copper stack in a 10\% ammonium persulfate (APS) solution for 1 to 1.5~hour at room temperature, without agitation, until the copper was fully removed. The quality and continuity of graphene after transfer were verified by optical microscopy and Raman spectroscopy.

\textbf{AFM characterization.} Atomic Force Microscopy (AFM) was carried out using either a Bruker Dimension Icon or a Asylum Research MFP-3D, in tapping mode (160AC probe, OPUS by MikroMasch) to measure topography and quantify the amplitude and periodicity of graphene folds.

\textbf{Raman spectroscopy.} Raman spectroscopy was performed (532 nm excitation, 100× objective) to map
strain-induced changes and evaluate defect signatures across the sample. Spectra were analyzed using Lorentzian fitting for the G, D, and 2D-bands.

\section*{Acknowledgements}

ASM thanks Didier Bégué (IPREM, UPPA) and Jacky Cresson (LMAP, UPPA) for very fruitful discussions which motivated the present work.
The authors thank S.Reynaud and N. Faure for technical support.
The authors acknowledge the financial support from CNRS MITI (Mission for Interdisciplinary and Transverse Initiatives) program, from Manutech-SLEIGHT (ANR 17-EURE-0026), and from ANR (French National Research Agency) with PRC projects 2D-PRESTO (ANR-19-CE09-0027) and SOFTNANOFLU (ANR-20-CE09-0025).

\bibstyle{achemso}
\bibliography{bibliography}

\providecommand{\latin}[1]{#1}
\makeatletter
\providecommand{\doi}
  {\begingroup\let\do\@makeother\dospecials
  \catcode`\{=1 \catcode`\}=2 \doi@aux}
\providecommand{\doi@aux}[1]{\endgroup\texttt{#1}}
\makeatother
\providecommand*\mcitethebibliography{\thebibliography}
\csname @ifundefined\endcsname{endmcitethebibliography}  {\let\endmcitethebibliography\endthebibliography}{}
\begin{mcitethebibliography}{51}
\providecommand*\natexlab[1]{#1}
\providecommand*\mciteSetBstSublistMode[1]{}
\providecommand*\mciteSetBstMaxWidthForm[2]{}
\providecommand*\mciteBstWouldAddEndPuncttrue
  {\def\EndOfBibitem{\unskip.}}
\providecommand*\mciteBstWouldAddEndPunctfalse
  {\let\EndOfBibitem\relax}
\providecommand*\mciteSetBstMidEndSepPunct[3]{}
\providecommand*\mciteSetBstSublistLabelBeginEnd[3]{}
\providecommand*\EndOfBibitem{}
\mciteSetBstSublistMode{f}
\mciteSetBstMaxWidthForm{subitem}{(\alph{mcitesubitemcount})}
\mciteSetBstSublistLabelBeginEnd
  {\mcitemaxwidthsubitemform\space}
  {\relax}
  {\relax}

\bibitem[Akinwande \latin{et~al.}(2017)Akinwande, Brennan, Bunch, Egberts, Felts, Gao, Huang, Kim, Li, Li, Liechti, Lu, Park, Reed, Wang, Yakobson, Zhang, Zhang, Zhou, and Zhu]{AKINWANDE201742}
Akinwande,~D. \latin{et~al.}  A review on mechanics and mechanical properties of 2D materials—Graphene and beyond. \emph{Extreme Mechanics Letters} \textbf{2017}, \emph{13}, 42--77\relax
\mciteBstWouldAddEndPuncttrue
\mciteSetBstMidEndSepPunct{\mcitedefaultmidpunct}
{\mcitedefaultendpunct}{\mcitedefaultseppunct}\relax
\EndOfBibitem
\bibitem[Sun \latin{et~al.}(2021)Sun, Papageorgiou, Humphreys, Dunstan, Puech, Proctor, Bousige, Machon, and San-Miguel]{sun2021APR}
Sun,~Y.~W.; Papageorgiou,~D.~G.; Humphreys,~C.~J.; Dunstan,~D.~J.; Puech,~P.; Proctor,~J.~E.; Bousige,~C.; Machon,~D.; San-Miguel,~A. Mechanical properties of graphene. \emph{Applied Physics Reviews} \textbf{2021}, \emph{8}, 021310\relax
\mciteBstWouldAddEndPuncttrue
\mciteSetBstMidEndSepPunct{\mcitedefaultmidpunct}
{\mcitedefaultendpunct}{\mcitedefaultseppunct}\relax
\EndOfBibitem
\bibitem[Galiotis \latin{et~al.}(2015)Galiotis, Frank, Koukaras, and Sfyris]{galiotis2015ARCBE}
Galiotis,~C.; Frank,~O.; Koukaras,~E.~N.; Sfyris,~D. Graphene Mechanics: Current Status and Perspectives. \emph{Annual Review of Chemical and Biomolecular Engineering} \textbf{2015}, \emph{6}, 121--140\relax
\mciteBstWouldAddEndPuncttrue
\mciteSetBstMidEndSepPunct{\mcitedefaultmidpunct}
{\mcitedefaultendpunct}{\mcitedefaultseppunct}\relax
\EndOfBibitem
\bibitem[Deng and Berry(2016)Deng, and Berry]{deng2016wrinkled}
Deng,~S.; Berry,~V. Wrinkled, rippled and crumpled graphene: an overview of formation mechanism, electronic properties, and applications. \emph{Materials Today} \textbf{2016}, \emph{19}, 197--212\relax
\mciteBstWouldAddEndPuncttrue
\mciteSetBstMidEndSepPunct{\mcitedefaultmidpunct}
{\mcitedefaultendpunct}{\mcitedefaultseppunct}\relax
\EndOfBibitem
\bibitem[Megra and Suk(2019)Megra, and Suk]{Megra_2019}
Megra,~Y.~T.; Suk,~J.~W. Adhesion properties of 2D materials. \emph{Journal of Physics D: Applied Physics} \textbf{2019}, \emph{52}, 364002\relax
\mciteBstWouldAddEndPuncttrue
\mciteSetBstMidEndSepPunct{\mcitedefaultmidpunct}
{\mcitedefaultendpunct}{\mcitedefaultseppunct}\relax
\EndOfBibitem
\bibitem[Lopez-Polin \latin{et~al.}(2022)Lopez-Polin, Gomez-Navarro, and Gomez-Herrero]{LOPEZPOLIN202218}
Lopez-Polin,~G.; Gomez-Navarro,~C.; Gomez-Herrero,~J. The effect of rippling on the mechanical properties of graphene. \emph{Nano Materials Science} \textbf{2022}, \emph{4}, 18--26, Special issue on Graphene and 2D Alternative Materials\relax
\mciteBstWouldAddEndPuncttrue
\mciteSetBstMidEndSepPunct{\mcitedefaultmidpunct}
{\mcitedefaultendpunct}{\mcitedefaultseppunct}\relax
\EndOfBibitem
\bibitem[Vandeparre \latin{et~al.}(2011)Vandeparre, Pi{\~n}eirua, Brau, Roman, Bico, Gay, Bao, Lau, Reis, and Damman]{vandeparre2010wrinkling}
Vandeparre,~H.; Pi{\~n}eirua,~M.; Brau,~F.; Roman,~B.; Bico,~J.; Gay,~C.; Bao,~W.; Lau,~C.~N.; Reis,~P.~M.; Damman,~P. Wrinkling hierarchy in constrained thin sheets from suspended graphene to curtains. \emph{Physical Review Letters} \textbf{2011}, \emph{106}, 224301\relax
\mciteBstWouldAddEndPuncttrue
\mciteSetBstMidEndSepPunct{\mcitedefaultmidpunct}
{\mcitedefaultendpunct}{\mcitedefaultseppunct}\relax
\EndOfBibitem
\bibitem[Zhang \latin{et~al.}(2011)Zhang, Akatyeva, and Dumitric\ifmmode~\u{a}\else \u{a}\fi{}]{ZhangPhysRevLett.106.255503}
Zhang,~D.-B.; Akatyeva,~E.; Dumitric\ifmmode~\u{a}\else \u{a}\fi{},~T. Bending Ultrathin Graphene at the Margins of Continuum Mechanics. \emph{Phys. Rev. Lett.} \textbf{2011}, \emph{106}, 255503\relax
\mciteBstWouldAddEndPuncttrue
\mciteSetBstMidEndSepPunct{\mcitedefaultmidpunct}
{\mcitedefaultendpunct}{\mcitedefaultseppunct}\relax
\EndOfBibitem
\bibitem[Zhu and et~al.(2012)Zhu, and et~al.]{Zhu2012nanolett}
Zhu,~W.; et~al. Structure and electronic transport in graphene wrinkles. \emph{Nano Letters} \textbf{2012}, \emph{12}, 3431--3436\relax
\mciteBstWouldAddEndPuncttrue
\mciteSetBstMidEndSepPunct{\mcitedefaultmidpunct}
{\mcitedefaultendpunct}{\mcitedefaultseppunct}\relax
\EndOfBibitem
\bibitem[Peng \latin{et~al.}(2020)Peng, Chen, Fan, Srolovitz, and Lei]{peng2020strain}
Peng,~Z.; Chen,~X.; Fan,~Y.; Srolovitz,~D.~J.; Lei,~D. Strain engineering of 2D semiconductors and graphene: from strain fields to band-structure tuning and photonic applications. \emph{Light: Science \& Applications} \textbf{2020}, \emph{9}, 190\relax
\mciteBstWouldAddEndPuncttrue
\mciteSetBstMidEndSepPunct{\mcitedefaultmidpunct}
{\mcitedefaultendpunct}{\mcitedefaultseppunct}\relax
\EndOfBibitem
\bibitem[Ni \latin{et~al.}(2008)Ni, Yu, Lu, Wang, Feng, and Shen]{ni2008uniaxial}
Ni,~Z.~H.; Yu,~T.; Lu,~Y.~H.; Wang,~Y.~Y.; Feng,~Y.~P.; Shen,~Z.~X. Uniaxial strain on graphene: Raman spectroscopy study and band-gap opening. \emph{ACS Nano} \textbf{2008}, \emph{2}, 2301--2305\relax
\mciteBstWouldAddEndPuncttrue
\mciteSetBstMidEndSepPunct{\mcitedefaultmidpunct}
{\mcitedefaultendpunct}{\mcitedefaultseppunct}\relax
\EndOfBibitem
\bibitem[Carrillo-Bastos \latin{et~al.}(2014)Carrillo-Bastos, Faria, Latg\'{e}, Mireles, and Sandler]{Carrillo-Bastos2014}
Carrillo-Bastos,~R.; Faria,~D.; Latg\'{e},~A.; Mireles,~F.; Sandler,~N. Strained fold-assisted transport in graphene nanoribbons. \emph{Physical Review B} \textbf{2014}, \emph{90}, 041411\relax
\mciteBstWouldAddEndPuncttrue
\mciteSetBstMidEndSepPunct{\mcitedefaultmidpunct}
{\mcitedefaultendpunct}{\mcitedefaultseppunct}\relax
\EndOfBibitem
\bibitem[Silva \latin{et~al.}(2015)Silva, Cresson, Rivaton, Begue, and Hiorns]{Silva2015}
Silva,~H.~S.; Cresson,~J.; Rivaton,~A.; Begue,~D.; Hiorns,~R.~C. Correlating geometry of multidimensional carbon allotropes molecules and stability. \emph{Organic Electronics} \textbf{2015}, \emph{26}, 395--399\relax
\mciteBstWouldAddEndPuncttrue
\mciteSetBstMidEndSepPunct{\mcitedefaultmidpunct}
{\mcitedefaultendpunct}{\mcitedefaultseppunct}\relax
\EndOfBibitem
\bibitem[Zang \latin{et~al.}(2013)Zang, Ryu, Pugno, Wang, Tu, Buehler, and Zhao]{zang2013multifunctionality}
Zang,~J.; Ryu,~S.; Pugno,~N.; Wang,~Q.; Tu,~Q.; Buehler,~M.~J.; Zhao,~X. Multifunctionality and control of the crumpling and unfolding of large-area graphene. \emph{Nature materials} \textbf{2013}, \emph{12}, 321--325\relax
\mciteBstWouldAddEndPuncttrue
\mciteSetBstMidEndSepPunct{\mcitedefaultmidpunct}
{\mcitedefaultendpunct}{\mcitedefaultseppunct}\relax
\EndOfBibitem
\bibitem[Bao \latin{et~al.}(2009)Bao, Miao, Z., Zhang, Dames, and Lau]{natBao2009contripp}
Bao,~W.; Miao,~F.; Z.,~C.; Zhang,~H.; Dames,~C.; Lau,~C. Controlled ripple texturing of suspended graphene and ultrathin graphite membranes. \emph{Nature nanotechnology} \textbf{2009}, \relax
\mciteBstWouldAddEndPunctfalse
\mciteSetBstMidEndSepPunct{\mcitedefaultmidpunct}
{}{\mcitedefaultseppunct}\relax
\EndOfBibitem
\bibitem[Tapaszt{\'o} \latin{et~al.}(2012)Tapaszt{\'o}, Dumitric{\u{a}}, Kim, Nemes-Incze, Hwang, and Bir{\'o}]{tapaszto2012breakdown}
Tapaszt{\'o},~L.; Dumitric{\u{a}},~T.; Kim,~S.~J.; Nemes-Incze,~P.; Hwang,~C.; Bir{\'o},~L.~P. Breakdown of continuum mechanics for nanometre-wavelength rippling of graphene. \emph{Nature physics} \textbf{2012}, \emph{8}, 739--742\relax
\mciteBstWouldAddEndPuncttrue
\mciteSetBstMidEndSepPunct{\mcitedefaultmidpunct}
{\mcitedefaultendpunct}{\mcitedefaultseppunct}\relax
\EndOfBibitem
\bibitem[Metten \latin{et~al.}(2014)Metten, Federspiel, Romeo, and Berciaud]{metten2014all}
Metten,~D.; Federspiel,~F.; Romeo,~M.; Berciaud,~S. All-optical blister test of suspended graphene using micro-Raman spectroscopy. \emph{Physical Review Applied} \textbf{2014}, \emph{2}, 054008\relax
\mciteBstWouldAddEndPuncttrue
\mciteSetBstMidEndSepPunct{\mcitedefaultmidpunct}
{\mcitedefaultendpunct}{\mcitedefaultseppunct}\relax
\EndOfBibitem
\bibitem[Amaral \latin{et~al.}(2021)Amaral, Forestier, Piednoir, Galafassi, Bousige, Machon, Pierre-Louis, Alencar, Souza~Filho, and San-Miguel]{amaral2021delamination}
Amaral,~I.; Forestier,~A.; Piednoir,~A.; Galafassi,~R.; Bousige,~C.; Machon,~D.; Pierre-Louis,~O.; Alencar,~R.; Souza~Filho,~A.; San-Miguel,~A. Delamination of multilayer graphene stacks from its substrate through wrinkle formation under high pressures. \emph{Carbon} \textbf{2021}, \emph{185}, 242--251\relax
\mciteBstWouldAddEndPuncttrue
\mciteSetBstMidEndSepPunct{\mcitedefaultmidpunct}
{\mcitedefaultendpunct}{\mcitedefaultseppunct}\relax
\EndOfBibitem
\bibitem[Wang \latin{et~al.}(2011)Wang, Yang, Shi, Zhang, Shi, Wang, and Zhang]{wang2011super}
Wang,~Y.; Yang,~R.; Shi,~Z.; Zhang,~L.; Shi,~D.; Wang,~E.; Zhang,~G. Super-elastic graphene ripples for flexible strain sensors. \emph{ACS nano} \textbf{2011}, \emph{5}, 3645--3650\relax
\mciteBstWouldAddEndPuncttrue
\mciteSetBstMidEndSepPunct{\mcitedefaultmidpunct}
{\mcitedefaultendpunct}{\mcitedefaultseppunct}\relax
\EndOfBibitem
\bibitem[Tsoukleri \latin{et~al.}(2009)Tsoukleri, Parthenios, Papagelis, Jalil, Ferrari, Geim, Novoselov, and Galiotis]{TsoukleriSMALL2009}
Tsoukleri,~G.; Parthenios,~J.; Papagelis,~K.; Jalil,~R.; Ferrari,~A.~C.; Geim,~A.~K.; Novoselov,~K.~S.; Galiotis,~C. Subjecting a Graphene Monolayer to Tension and Compression. \emph{Small} \textbf{2009}, \emph{5}, 2397--2402\relax
\mciteBstWouldAddEndPuncttrue
\mciteSetBstMidEndSepPunct{\mcitedefaultmidpunct}
{\mcitedefaultendpunct}{\mcitedefaultseppunct}\relax
\EndOfBibitem
\bibitem[Frank \latin{et~al.}(2010)Frank, Tsoukleri, Parthenios, Papagelis, Riaz, Jalil, Novoselov, and Galiotis]{FrankACS2010}
Frank,~O.; Tsoukleri,~G.; Parthenios,~J.; Papagelis,~K.; Riaz,~I.; Jalil,~R.; Novoselov,~K.~S.; Galiotis,~C. Compression Behavior of Single-Layer Graphenes. \emph{ACS Nano} \textbf{2010}, \emph{4}, 3131--3138, PMID: 20496881\relax
\mciteBstWouldAddEndPuncttrue
\mciteSetBstMidEndSepPunct{\mcitedefaultmidpunct}
{\mcitedefaultendpunct}{\mcitedefaultseppunct}\relax
\EndOfBibitem
\bibitem[Jiang \latin{et~al.}(2014)Jiang, Huang, and Zhu]{jiang2014interfacial}
Jiang,~T.; Huang,~R.; Zhu,~Y. Interfacial sliding and buckling of monolayer graphene on a stretchable substrate. \emph{Advanced Functional Materials} \textbf{2014}, \emph{24}, 396--402\relax
\mciteBstWouldAddEndPuncttrue
\mciteSetBstMidEndSepPunct{\mcitedefaultmidpunct}
{\mcitedefaultendpunct}{\mcitedefaultseppunct}\relax
\EndOfBibitem
\bibitem[Leem \latin{et~al.}(2019)Leem, Lee, Wang, Kim, Mun, Haque, Kang, and Nam]{Leem_2019}
Leem,~J.; Lee,~Y.; Wang,~M.~C.; Kim,~J.~M.; Mun,~J.; Haque,~M.~F.; Kang,~S.-W.; Nam,~S. Crack-assisted, localized deformation of van der Waals materials for enhanced strain confinement. \emph{2D Materials} \textbf{2019}, \emph{6}, 044001\relax
\mciteBstWouldAddEndPuncttrue
\mciteSetBstMidEndSepPunct{\mcitedefaultmidpunct}
{\mcitedefaultendpunct}{\mcitedefaultseppunct}\relax
\EndOfBibitem
\bibitem[Hallam \latin{et~al.}(2015)Hallam, Shakouri, Poliani, Rooney, Ivanov, Potie, Taylor, Bonn, Turchinovich, Haigh, \latin{et~al.} others]{hallam2015controlled}
Hallam,~T.; Shakouri,~A.; Poliani,~E.; Rooney,~A.~P.; Ivanov,~I.; Potie,~A.; Taylor,~H.~K.; Bonn,~M.; Turchinovich,~D.; Haigh,~S.~J.; others Controlled folding of graphene: GraFold printing. \emph{Nano letters} \textbf{2015}, \emph{15}, 857--863\relax
\mciteBstWouldAddEndPuncttrue
\mciteSetBstMidEndSepPunct{\mcitedefaultmidpunct}
{\mcitedefaultendpunct}{\mcitedefaultseppunct}\relax
\EndOfBibitem
\bibitem[Reserbat-Plantey \latin{et~al.}(2014)Reserbat-Plantey, Kalita, Han, Ferlazzo, Autier-Laurent, Komatsu, Li, Weil, Ralko, Marty, \latin{et~al.} others]{reserbat2014strain}
Reserbat-Plantey,~A.; Kalita,~D.; Han,~Z.; Ferlazzo,~L.; Autier-Laurent,~S.; Komatsu,~K.; Li,~C.; Weil,~R.; Ralko,~A.; Marty,~L.; others Strain superlattices and macroscale suspension of graphene induced by corrugated substrates. \emph{Nano letters} \textbf{2014}, \emph{14}, 5044--5051\relax
\mciteBstWouldAddEndPuncttrue
\mciteSetBstMidEndSepPunct{\mcitedefaultmidpunct}
{\mcitedefaultendpunct}{\mcitedefaultseppunct}\relax
\EndOfBibitem
\bibitem[Pacakova \latin{et~al.}(2017)Pacakova, Verhagen, Bousa, H{\"u}bner, Vejpravova, Kalbac, and Frank]{pacakova2017mastering}
Pacakova,~B.; Verhagen,~T.; Bousa,~M.; H{\"u}bner,~U.; Vejpravova,~J.; Kalbac,~M.; Frank,~O. Mastering the wrinkling of self-supported graphene. \emph{Scientific reports} \textbf{2017}, \emph{7}, 10003\relax
\mciteBstWouldAddEndPuncttrue
\mciteSetBstMidEndSepPunct{\mcitedefaultmidpunct}
{\mcitedefaultendpunct}{\mcitedefaultseppunct}\relax
\EndOfBibitem
\bibitem[Chen \latin{et~al.}(2019)Chen, Zhang, Zhang, Wang, Bao, Que, Xiao, Du, Ouyang, Pantelides, \latin{et~al.} others]{chen2019atomically}
Chen,~H.; Zhang,~X.-L.; Zhang,~Y.-Y.; Wang,~D.; Bao,~D.-L.; Que,~Y.; Xiao,~W.; Du,~S.; Ouyang,~M.; Pantelides,~S.~T.; others Atomically precise, custom-design origami graphene nanostructures. \emph{Science} \textbf{2019}, \emph{365}, 1036--1040\relax
\mciteBstWouldAddEndPuncttrue
\mciteSetBstMidEndSepPunct{\mcitedefaultmidpunct}
{\mcitedefaultendpunct}{\mcitedefaultseppunct}\relax
\EndOfBibitem
\bibitem[Chang \latin{et~al.}(2018)Chang, Kim, Sung, Yeon, Chang, Li, and Kim]{chang2018small}
Chang,~J.~S.; Kim,~S.; Sung,~H.-J.; Yeon,~J.; Chang,~K.~J.; Li,~X.; Kim,~S. Graphene Nanoribbons with Atomically Sharp Edges Produced by AFM Induced Self-Folding. \emph{Small} \textbf{2018}, \emph{14}, 1803386\relax
\mciteBstWouldAddEndPuncttrue
\mciteSetBstMidEndSepPunct{\mcitedefaultmidpunct}
{\mcitedefaultendpunct}{\mcitedefaultseppunct}\relax
\EndOfBibitem
\bibitem[Blees \latin{et~al.}(2015)Blees, Barnard, Rose, Roberts, McGill, Huang, Ruyack, Kevek, Kobrin, Muller, and McEuen]{blees2015graphenekiri}
Blees,~M.~K.; Barnard,~A.~W.; Rose,~P.~A.; Roberts,~S.~P.; McGill,~K.~L.; Huang,~P.~Y.; Ruyack,~A.~R.; Kevek,~J.~W.; Kobrin,~B.; Muller,~D.~A.; McEuen,~P.~L. Graphene kirigami. \emph{Nature} \textbf{2015}, \emph{524}, 204--207\relax
\mciteBstWouldAddEndPuncttrue
\mciteSetBstMidEndSepPunct{\mcitedefaultmidpunct}
{\mcitedefaultendpunct}{\mcitedefaultseppunct}\relax
\EndOfBibitem
\bibitem[Zhang \latin{et~al.}(2021)Zhang, Tian, Mei, and Di]{ZHANG20211MSERR}
Zhang,~Z.; Tian,~Z.; Mei,~Y.; Di,~Z. Shaping and structuring 2D materials via kirigami and origami. \emph{Materials Science and Engineering: R: Reports} \textbf{2021}, \emph{145}, 100621\relax
\mciteBstWouldAddEndPuncttrue
\mciteSetBstMidEndSepPunct{\mcitedefaultmidpunct}
{\mcitedefaultendpunct}{\mcitedefaultseppunct}\relax
\EndOfBibitem
\bibitem[Khonina \latin{et~al.}(2020)Khonina, Kazanskiy, Karpeev, and Butt]{khonina2020bessel}
Khonina,~S.~N.; Kazanskiy,~N.~L.; Karpeev,~S.~V.; Butt,~M.~A. Bessel beam: Significance and applications—A progressive review. \emph{Micromachines} \textbf{2020}, \emph{11}, 997\relax
\mciteBstWouldAddEndPuncttrue
\mciteSetBstMidEndSepPunct{\mcitedefaultmidpunct}
{\mcitedefaultendpunct}{\mcitedefaultseppunct}\relax
\EndOfBibitem
\bibitem[Bhuyan \latin{et~al.}(2010)Bhuyan, Courvoisier, Lacourt, Jacquot, Furfaro, Withford, and Dudley]{bhuyan2010ultrafast}
Bhuyan,~M.; Courvoisier,~F.; Lacourt,~P.-A.; Jacquot,~M.; Furfaro,~L.; Withford,~M.; Dudley,~J. Ultrafast Bessel beams for high aspect ratio taper free micromachining of glass. Nonlinear Optics and Applications IV. 2010; pp 341--348\relax
\mciteBstWouldAddEndPuncttrue
\mciteSetBstMidEndSepPunct{\mcitedefaultmidpunct}
{\mcitedefaultendpunct}{\mcitedefaultseppunct}\relax
\EndOfBibitem
\bibitem[Mao \latin{et~al.}(2004)Mao, Qu{\'e}r{\'e}, Guizard, Mao, Russo, Petite, and Martin]{mao2004dynamics}
Mao,~S.; Qu{\'e}r{\'e},~F.; Guizard,~S.; Mao,~X.; Russo,~R.; Petite,~G.; Martin,~P. Dynamics of femtosecond laser interactions with dielectrics. \emph{Applied Physics A} \textbf{2004}, \emph{79}, 1695--1709\relax
\mciteBstWouldAddEndPuncttrue
\mciteSetBstMidEndSepPunct{\mcitedefaultmidpunct}
{\mcitedefaultendpunct}{\mcitedefaultseppunct}\relax
\EndOfBibitem
\bibitem[Gattass and Mazur(2008)Gattass, and Mazur]{gattass2008femtosecond}
Gattass,~R.~R.; Mazur,~E. Femtosecond laser micromachining in transparent materials. \emph{Nature photonics} \textbf{2008}, \emph{2}, 219--225\relax
\mciteBstWouldAddEndPuncttrue
\mciteSetBstMidEndSepPunct{\mcitedefaultmidpunct}
{\mcitedefaultendpunct}{\mcitedefaultseppunct}\relax
\EndOfBibitem
\bibitem[Rebollar \latin{et~al.}(2011)Rebollar, P{\'e}rez, Hern{\'a}ndez, Mart{\'\i}n-Fabiani, Rueda, Ezquerra, and Castillejo]{rebollar2011assessment}
Rebollar,~E.; P{\'e}rez,~S.; Hern{\'a}ndez,~J.~J.; Mart{\'\i}n-Fabiani,~I.; Rueda,~D.~R.; Ezquerra,~T.~A.; Castillejo,~M. Assessment and formation mechanism of laser-induced periodic surface structures on polymer spin-coated films in real and reciprocal space. \emph{Langmuir} \textbf{2011}, \emph{27}, 5596--5606\relax
\mciteBstWouldAddEndPuncttrue
\mciteSetBstMidEndSepPunct{\mcitedefaultmidpunct}
{\mcitedefaultendpunct}{\mcitedefaultseppunct}\relax
\EndOfBibitem
\bibitem[Rebollar \latin{et~al.}(2015)Rebollar, Castillejo, and Ezquerra]{rebollar2015laser}
Rebollar,~E.; Castillejo,~M.; Ezquerra,~T.~A. Laser induced periodic surface structures on polymer films: From fundamentals to applications. \emph{European Polymer Journal} \textbf{2015}, \emph{73}, 162--174\relax
\mciteBstWouldAddEndPuncttrue
\mciteSetBstMidEndSepPunct{\mcitedefaultmidpunct}
{\mcitedefaultendpunct}{\mcitedefaultseppunct}\relax
\EndOfBibitem
\bibitem[Stankova \latin{et~al.}(2016)Stankova, Atanasov, Nikov, Nikov, Nedyalkov, Stoyanchov, Fukata, Kolev, Valova, Georgieva, \latin{et~al.} others]{stankova2016optical}
Stankova,~N.; Atanasov,~P.; Nikov,~R.~G.; Nikov,~R.; Nedyalkov,~N.; Stoyanchov,~T.; Fukata,~N.; Kolev,~K.; Valova,~E.; Georgieva,~J.; others Optical properties of polydimethylsiloxane (PDMS) during nanosecond laser processing. \emph{Applied Surface Science} \textbf{2016}, \emph{374}, 96--103\relax
\mciteBstWouldAddEndPuncttrue
\mciteSetBstMidEndSepPunct{\mcitedefaultmidpunct}
{\mcitedefaultendpunct}{\mcitedefaultseppunct}\relax
\EndOfBibitem
\bibitem[Han \latin{et~al.}(2014)Han, Kimouche, Kalita, Allain, Arjmandi-Tash, Reserbat-Plantey, Marty, Pairis, Reita, Bendiab, Coraux, and Bouchiat]{HanCVDafm2014}
Han,~Z.; Kimouche,~A.; Kalita,~D.; Allain,~A.; Arjmandi-Tash,~H.; Reserbat-Plantey,~A.; Marty,~L.; Pairis,~S.; Reita,~V.; Bendiab,~N.; Coraux,~J.; Bouchiat,~V. Homogeneous Optical and Electronic Properties of Graphene Due to the Suppression of Multilayer Patches During CVD on Copper Foils. \emph{Advanced Functional Materials} \textbf{2014}, \emph{24}, 964--970\relax
\mciteBstWouldAddEndPuncttrue
\mciteSetBstMidEndSepPunct{\mcitedefaultmidpunct}
{\mcitedefaultendpunct}{\mcitedefaultseppunct}\relax
\EndOfBibitem
\bibitem[Coraux \latin{et~al.}(2013)Coraux, Marty, Bendiab, and Bouchiat]{coraux2013functional}
Coraux,~J.; Marty,~L.; Bendiab,~N.; Bouchiat,~V. Functional hybrid systems based on large-area high-quality graphene. \emph{Accounts of Chemical Research} \textbf{2013}, \emph{46}, 2193--2201\relax
\mciteBstWouldAddEndPuncttrue
\mciteSetBstMidEndSepPunct{\mcitedefaultmidpunct}
{\mcitedefaultendpunct}{\mcitedefaultseppunct}\relax
\EndOfBibitem
\bibitem[Arjmandi-Tash \latin{et~al.}(2018)Arjmandi-Tash, Kalita, Han, Othmen, Nayak, Berne, Landers, Watanabe, Taniguchi, Marty, \latin{et~al.} others]{arjmandi2018large}
Arjmandi-Tash,~H.; Kalita,~D.; Han,~Z.; Othmen,~R.; Nayak,~G.; Berne,~C.; Landers,~J.; Watanabe,~K.; Taniguchi,~T.; Marty,~L.; others Large scale graphene/h-BN heterostructures obtained by direct CVD growth of graphene using high-yield proximity-catalytic process. \emph{Journal of Physics: Materials} \textbf{2018}, \emph{1}, 015003\relax
\mciteBstWouldAddEndPuncttrue
\mciteSetBstMidEndSepPunct{\mcitedefaultmidpunct}
{\mcitedefaultendpunct}{\mcitedefaultseppunct}\relax
\EndOfBibitem
\bibitem[Ferrari \latin{et~al.}(2006)Ferrari, Meyer, Scardaci, Casiraghi, Lazzeri, Mauri, Piscanec, Jiang, Novoselov, Roth, and Geim]{ferrari2006raman}
Ferrari,~A.~C.; Meyer,~J.~C.; Scardaci,~V.; Casiraghi,~C.; Lazzeri,~M.; Mauri,~F.; Piscanec,~S.; Jiang,~D.; Novoselov,~K.~S.; Roth,~S.; Geim,~A.~K. Raman spectroscopy of graphene and graphite: Disorder, electron--phonon coupling, doping and nonadiabatic effects. \emph{Physical Review Letters} \textbf{2006}, \emph{97}, 187401\relax
\mciteBstWouldAddEndPuncttrue
\mciteSetBstMidEndSepPunct{\mcitedefaultmidpunct}
{\mcitedefaultendpunct}{\mcitedefaultseppunct}\relax
\EndOfBibitem
\bibitem[Malard \latin{et~al.}(2009)Malard, Pimenta, Dresselhaus, and Dresselhaus]{malard2009raman}
Malard,~L.~M.; Pimenta,~M.~A.; Dresselhaus,~G.; Dresselhaus,~M.~S. Raman spectroscopy in graphene. \emph{Physics Reports} \textbf{2009}, \emph{473}, 51--87\relax
\mciteBstWouldAddEndPuncttrue
\mciteSetBstMidEndSepPunct{\mcitedefaultmidpunct}
{\mcitedefaultendpunct}{\mcitedefaultseppunct}\relax
\EndOfBibitem
\bibitem[Yamamoto \latin{et~al.}(2012)Yamamoto, Pierre-Louis, Huang, Fuhrer, Einstein, and Cullen]{yamamoto2012princess}
Yamamoto,~M.; Pierre-Louis,~O.; Huang,~J.; Fuhrer,~M.~S.; Einstein,~T.~L.; Cullen,~W.~G. “The Princess and the Pea” at the Nanoscale: Wrinkling and Delamination of Graphene on Nanoparticles. \emph{Physical Review X} \textbf{2012}, \emph{2}, 041018\relax
\mciteBstWouldAddEndPuncttrue
\mciteSetBstMidEndSepPunct{\mcitedefaultmidpunct}
{\mcitedefaultendpunct}{\mcitedefaultseppunct}\relax
\EndOfBibitem
\bibitem[Vella \latin{et~al.}(2009)Vella, Bico, Boudaoud, and Reis]{vella2009macroscopic}
Vella,~D.; Bico,~J.; Boudaoud,~A.; Reis,~P.~M. The macroscopic delamination of thin films from elastic substrates. \emph{Proceedings of the National Academy of Sciences} \textbf{2009}, \emph{106}, 10901--10906\relax
\mciteBstWouldAddEndPuncttrue
\mciteSetBstMidEndSepPunct{\mcitedefaultmidpunct}
{\mcitedefaultendpunct}{\mcitedefaultseppunct}\relax
\EndOfBibitem
\bibitem[Brennan \latin{et~al.}(2015)Brennan, Nguyen, Yu, and Lu]{brennanAMI2015}
Brennan,~C.~J.; Nguyen,~J.; Yu,~E.~T.; Lu,~N. Interface Adhesion between 2D Materials and Elastomers Measured by Buckle Delaminations. \emph{Advanced Materials Interfaces} \textbf{2015}, \emph{2}, 1500176\relax
\mciteBstWouldAddEndPuncttrue
\mciteSetBstMidEndSepPunct{\mcitedefaultmidpunct}
{\mcitedefaultendpunct}{\mcitedefaultseppunct}\relax
\EndOfBibitem
\bibitem[Picheau \latin{et~al.}(2021)Picheau, Impellizzeri, Rybkovskiy, Bayle, Mevellec, Hof, Saadaoui, No{\'e}, Dias~Torres, Duvail, Monthioux, Humbert, Puech, Ewels, and P{\'e}nicaud]{picheau2021intense}
Picheau,~E.; Impellizzeri,~A.; Rybkovskiy,~D.; Bayle,~M.; Mevellec,~J.-Y.; Hof,~F.; Saadaoui,~H.; No{\'e},~L.; Dias~Torres,~A.; Duvail,~J.-L.; Monthioux,~M.; Humbert,~B.; Puech,~P.; Ewels,~C.~P.; P{\'e}nicaud,~A. Intense Raman D Band without Disorder in Flattened Carbon Nanotubes. \emph{ACS Nano} \textbf{2021}, \emph{15}, 596--603\relax
\mciteBstWouldAddEndPuncttrue
\mciteSetBstMidEndSepPunct{\mcitedefaultmidpunct}
{\mcitedefaultendpunct}{\mcitedefaultseppunct}\relax
\EndOfBibitem
\bibitem[Galafassi \latin{et~al.}(2024)Galafassi, Vialla, Pischedda, Diaf, and San-Miguel]{GALAFASSIcarbon2024}
Galafassi,~R.; Vialla,~F.; Pischedda,~V.; Diaf,~H.; San-Miguel,~A. Reversible Raman D-band changes: A new probe into the pressure-induced collapse of carbon nanotubes. \emph{Carbon} \textbf{2024}, \emph{229}, 119447\relax
\mciteBstWouldAddEndPuncttrue
\mciteSetBstMidEndSepPunct{\mcitedefaultmidpunct}
{\mcitedefaultendpunct}{\mcitedefaultseppunct}\relax
\EndOfBibitem
\bibitem[Kim \latin{et~al.}(2019)Kim, Leem, Kim, Kang, Choi, Haque, Kang, and Nam]{kim2019uniaxially}
Kim,~J.; Leem,~J.; Kim,~H.~N.; Kang,~P.; Choi,~J.; Haque,~M.~F.; Kang,~D.; Nam,~S. Uniaxially crumpled graphene as a platform for guided myotube formation. \emph{Microsystems \& nanoengineering} \textbf{2019}, \emph{5}, 53\relax
\mciteBstWouldAddEndPuncttrue
\mciteSetBstMidEndSepPunct{\mcitedefaultmidpunct}
{\mcitedefaultendpunct}{\mcitedefaultseppunct}\relax
\EndOfBibitem
\bibitem[Castellanos-Gomez \latin{et~al.}(2013)Castellanos-Gomez, Rold{\'a}n, Cappelluti, Buscema, Guinea, Van Der~Zant, and Steele]{castellanos2013local}
Castellanos-Gomez,~A.; Rold{\'a}n,~R.; Cappelluti,~E.; Buscema,~M.; Guinea,~F.; Van Der~Zant,~H.~S.; Steele,~G.~A. Local strain engineering in atomically thin MoS2. \emph{Nano letters} \textbf{2013}, \emph{13}, 5361--5366\relax
\mciteBstWouldAddEndPuncttrue
\mciteSetBstMidEndSepPunct{\mcitedefaultmidpunct}
{\mcitedefaultendpunct}{\mcitedefaultseppunct}\relax
\EndOfBibitem
\bibitem[Kumar \latin{et~al.}(2015)Kumar, Kaczmarczyk, and Gerardot]{kumar2015strain}
Kumar,~S.; Kaczmarczyk,~A.; Gerardot,~B.~D. Strain-induced spatial and spectral isolation of quantum emitters in mono-and bilayer WSe2. \emph{Nano letters} \textbf{2015}, \emph{15}, 7567--7573\relax
\mciteBstWouldAddEndPuncttrue
\mciteSetBstMidEndSepPunct{\mcitedefaultmidpunct}
{\mcitedefaultendpunct}{\mcitedefaultseppunct}\relax
\EndOfBibitem
\end{mcitethebibliography}

\includepdf[pages=-]{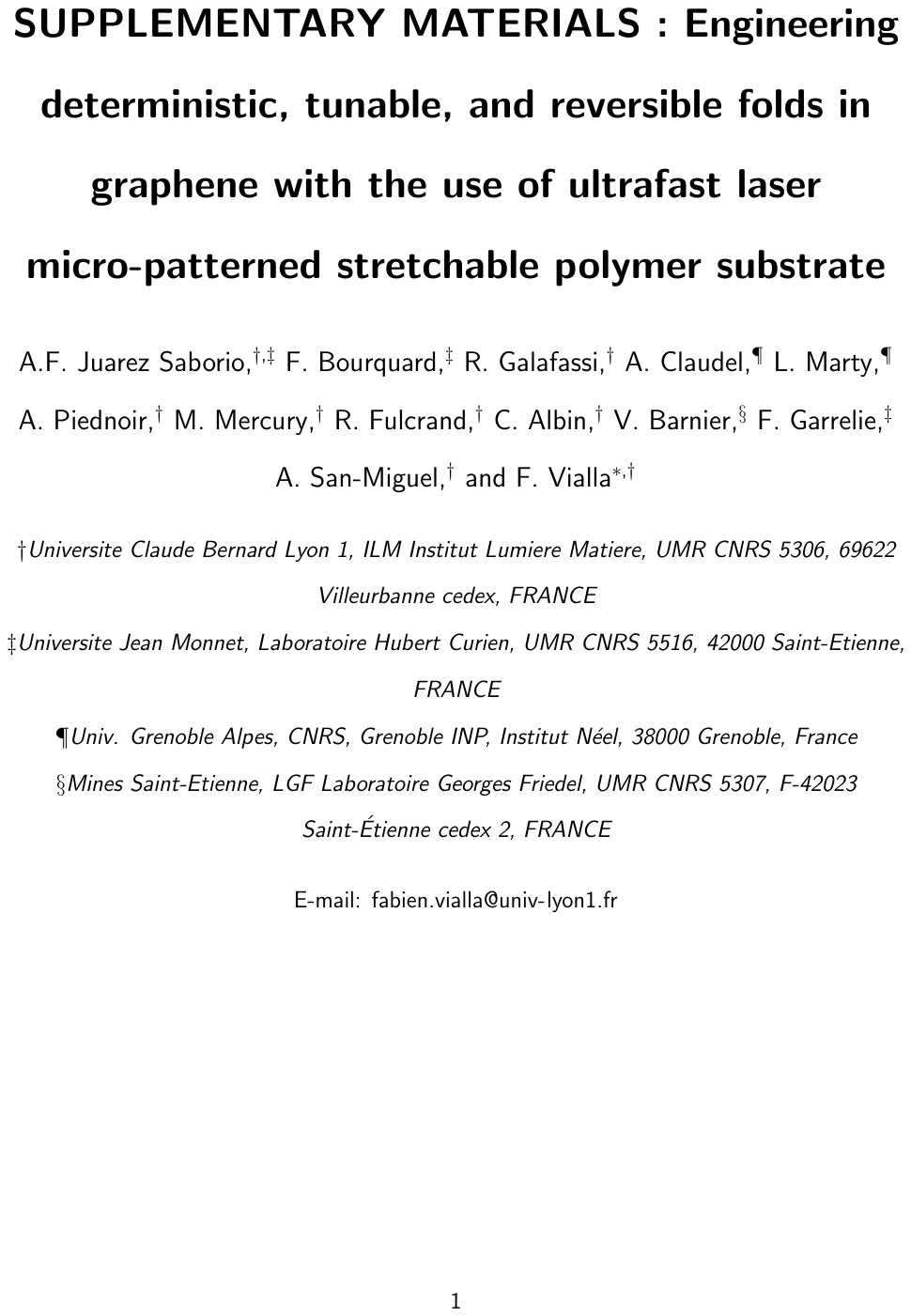}

\end{document}